\documentclass[letterpaper, 10 pt, conference]{ieeeconf}  % Comment this line out if you need a4paper

\IEEEoverridecommandlockouts                              % This command is only needed if 
\usepackage{hyperref} 
\usepackage{mathrsfs}
\usepackage[font={small}]{caption}
\usepackage{scalerel}
\usepackage{url}
\usepackage{bbold}
\usepackage{amsfonts}
\usepackage{amsmath,amssymb,amsfonts}
\usepackage{graphicx}
\usepackage{hyperref}
\usepackage{wrapfig}
\usepackage{mathrsfs} 
\usepackage{algorithm,algorithmic}
\usepackage{array}
\usepackage{times}
\usepackage{url}
\usepackage{subfigure}
\usepackage{cite}
\usepackage{upgreek}
\usepackage{float}
\usepackage{longtable}
\usepackage{color}
\usepackage{wasysym}
\usepackage{grffile}

\allowdisplaybreaks

\newcommand\numberthis{\addtocounter{equation}{1}\tag{\theequation}}

\newcommand{\0}{\mathbb{0}}
\newcommand{\1}{\mathbb{1}}

\newcommand{\E}{\mathbb{E}}
\newcommand{\R}{\mathbb{R}}

\newcommand{\ab}{\mathbf{a}}
\newcommand{\eb}{\mathbf{e}}

\newcommand{\vb}{\mathbf{v}}

\newcommand{\xb}{\mathbf{x}}
\newcommand{\yb}{\mathbf{y}}

\newcommand{\Ab}{\mathbf{A}}
\newcommand{\Bb}{\mathbf{B}}

\newcommand{\Eb}{\mathbf{E}}

\newcommand{\Gb}{\mathbf{G}}
\newcommand{\Hb}{\mathbf{H}}
\newcommand{\Ib}{\mathbf{I}}

\newcommand{\Lb}{\mathbf{L}}
\newcommand{\Pb}{\mathbf{P}}
\newcommand{\Qb}{\mathbf{Q}}

\newcommand{\omegab}{\boldsymbol{\omega}}

\newcommand{\nub}{\boldsymbol{\nu}}
\newcommand{\thetab}{\boldsymbol{\theta}}

\newcommand{\Lc}{\mathcal{L}}
\newcommand{\Nc}{\mathcal{N}}

\newcommand{\Uc}{\mathcal{U}}
\newcommand{\Xc}{\mathcal{X}}
\newcommand{\Yc}{\mathcal{Y}}

\newcommand{\fh}{\widehat{f}}

\newcommand{\norm}[1]{\left\lVert#1\right\rVert}
\newcommand{\tr}[1]{\text{Tr}\left[#1\right]}

\newcommand{\ex}[1]{\E\left[#1\right]}

\newtheorem{theorem}{Theorem}

\newtheorem{assumption}{Assumption}

\newtheorem{proposition}[theorem]{Proposition}

\title{\LARGE \bf Distributed Parameter Estimation in Randomized\\ One-hidden-layer Neural Networks}

\author{Yinsong Wang and Shahin Shahrampour 
\thanks{Yinsong Wang and Shahin Shahrampour are with the Department of Industrial and Systems Engineering at Texas A\&M University, College Station, TX 77843, USA. 
        {\tt\small email:\{gritti@tamu.edu;shahin@tamu.edu\}}.}%
}

\begin{document}

\maketitle
\thispagestyle{empty}
\pagestyle{empty}

%%%%%%%%%%%%%%%%%%%%%%%%%%%%%%%%%%%%%%%%%%%%%%%%%%%%%%%%%%%%%%%%%%%%%%%%%%%%%%%%
\begin{abstract}
This paper addresses distributed parameter estimation in randomized one-hidden-layer neural networks. A group of agents sequentially receive measurements of an unknown parameter that is only partially observable to them. In this paper, we present a fully distributed estimation algorithm where agents exchange local estimates with their neighbors to collectively identify the true value of the parameter. We prove that this distributed update provides an asymptotically unbiased estimator of the unknown parameter, i.e., the first moment of the expected global error converges to zero asymptotically. We further analyze the efficiency of the proposed estimation scheme by establishing an asymptotic upper bound on the variance of the global error. Applying our method to a real-world dataset related to appliances energy prediction, we observe that our empirical findings verify the theoretical results.
\end{abstract}

%%%%%%%%%%%%%%%%%%%%%%%%%%%%%%%%%%%%%%%%%%%%%%%%%%%%%%%%%%%%%%%%%%%%%%%%%%%%%%%%
\section{Introduction}
Supervised learning is a fundamental machine learning problem, where given input-output data samples, a learner aims to find a mapping (or function) from inputs to outputs \cite{friedman2001elements}. A good mapping is one that can be used for prediction of outputs corresponding to previously unseen inputs. Recently, deep neural networks have dominated the task of supervised learning in various applications,  including computer vision \cite{he2016deep}, speech recognition \cite{amodei2016deep}, robotics \cite{levine+2016}, and biomedical image analysis \cite{shen2017deep}. These methods, however, are data hungry and their application to domains with few/sparse labeled samples remains an active field of research \cite{noroozi2018boosting}. An alternative effective method for supervised learning is shallow architectures with one-hidden-layer. This architecture was motivated by the classical results of Cybenko \cite{cybenko1989approximation} and Barron \cite{barron1993universal}, showing that (under some technical assumptions) one can use sigmoidal basis functions to approximate any output that is a continuous function of the input. These results later motivated researchers to develop algorithmic frameworks to leverage shallow networks for data representation. The seminal work of Rahimi and Recht is a prominent point in case \cite{rahimi2009weighted}. In their approach, the nonlinear basis functions are selected using Monte-Carlo sampling with a theoretical guarantee that the approximated function converges asymptotically with respect to the number of data samples and basis functions.

The problem of function approximation in supervised learning (both in shallow and deep neural networks) is often formulated via {\it empirical risk minimization} \cite{friedman2001elements}, which amounts to solving an optimization problem over a high-dimensional parameter. Due to the computational challenges associated with high-dimensional optimization, an appealing solution turns out to be decentralized training of neural networks \cite{mcmahan2017communication}. On the other hand, recent advancement in distributed computing within control and signal processing communities \cite{khan2010connectivity,stankovic2011decentralized,kar2012distributed,shahrampour2013exponentially,atanasov2014joint,mitra2016approach} has provided novel decentralized techniques for parameter estimation over multi-agent networks. In these scenarios, each individual agent receives partially informative measurements about the parameter and engages in local communications with other agents to collaboratively accomplish the global task. A crucial component of these methods is a {\it consensus} protocol \cite{jadbabaie2003coordination}, allowing collective information aggregation and estimation. Distributed algorithms gained popularity due to their ability to handle large data sets, low computational burden over agents, and robustness to failure of a central agent.  

Motivated by the importance of distributed computing in high-dimensional parameter estimation, in this paper, we consider distributed parameter estimation in randomized one-hidden-layer neural networks. A group of agents sequentially obtain low-dimensional measurements of the parameter (in various locations at different randomized frequencies). Despite the parameter being partially observable to each individual agent, the global spread of measurements is informative enough for a collective estimation. We propose a fully distributed update where each agent engages in local interactions with its neighboring agents to construct iterative estimates of the parameter. The update is akin to {\it consensus+innovation} algorithms in the distributed estimation literature \cite{khan2010connectivity,kar2012distributed,shahrampour2016distributed}.

Our main {\it theoretical} contribution is to characterize the first and second moments of the global estimation error. In particular, we prove that the distributed update provides an asymptotically unbiased estimator of the unknown parameter when the randomness of data samples is expected out, i.e., the first moment of the global error converges to zero asymptotically. This result also allows us to characterize the convergence rate and derive a feasible range for innovation rate. We further analyze the efficiency of the proposed estimation scheme by establishing an asymptotic upper bound on the second moment of the global error. We finally simulate our method on a real-world data related to appliances energy prediction, where we observe that our empirical findings verify the theoretical results.

\section{Problem Statement}
\noindent
{\bf Notation:}
We adhere to the following notation table throughout the paper:

\begin{center}
    \begin{tabular}{|c||c|}
    \hline
         $[n]$& set $\{1,2,3,...,n\}$ for any integer $n$  \\
         \hline
         $\xb^\top$ & transpose of vector $\xb$ \\
         \hline
         $\Ib_M$ & identity matrix of size $M$\\
         \hline
         $\1_n$ & vector of all ones with dimension $n$\\
         \hline
         $\0$ & vector of all zeros\\
         \hline
         $\norm{\cdot}_p$ & $\Lc_p$-norm operator\\
         \hline
         $\lambda_i(\Pb)$ & $i$-th largest eigenvalue of matrix $\Pb$\\
         \hline
         $\E[\cdot]$ & expectation operator\\
         \hline
         $\rho(\Qb)$ & spectral radius of matrix $\Qb$\\
         \hline
         $\tr{\cdot}$ & trace operator\\
         \hline
         $\Ab\preceq\Bb$ & $\Bb-\Ab$ is positive semi-definite\\
    \hline     
    \end{tabular}
\end{center}
The vectors are in column format. Boldface lowercase variables (e.g., $\ab$) are used for vectors, and boldface uppercase variables (e.g., $\Ab$) are used for matrices.

\subsection{One-Hidden-Layer Neural Networks: The Centralized Problem}\label{2A}
Let us consider a regression problem of the form
\begin{equation*}
    y = f(\xb) + v,
\end{equation*}
where $y \in \Yc \subseteq \R$ is the output, $\xb \in \Xc \subseteq \R^d$ is the input, and $v$ is the noise term with zero mean and constant variance. The objective is to find the {\it unknown} mapping (or function) $f: \Xc \to \Yc$ based on available input-output pairs $\{(\xb_j,y_j)\}$. Various regression methods assume different functional forms to approximate $f(\cdot)$. For example, in linear regression, the input-output relationship is assumed to follow a linear model. In this work, we focus on one-hidden-layer neural networks \cite{cybenko1989approximation}, where the approximated function $\fh(\cdot)$ is a nonlinear function of the input, and

\begin{equation}\label{OHL}
    \fh(\xb) = \sum_{l=1}^M \theta_l\phi(\xb,\omegab_l),
\end{equation}
where $\phi$ is called a basis function (or \textit{feature map}) parameterized by $\omegab_l$. In the above model, the parameters $\omegab_l$ and $\theta_l$ are unknown and should be learned from data (i.e., input-output pairs). The underlying intuition behind this model is that the feature map transforms the original data from dimension $d$ to $M$, where often time we have $M\gg d$. Since the new space has a higher dimension, it provides more flexibility for approximation of the unknown function (as opposed to a linear model that is restrictive). It turns out that approximations of form \eqref{OHL} are dense in the space of continuous functions \cite{cybenko1989approximation}, i.e., they can be used to approximate any continuous function (on the unit cube). 

However, from an algorithmic perspective, learning both $\theta_l$ and $\omegab_l$ is computationally expensive. For a nonlinear feature map $\phi$ (e.g., cosine feature map), the problem is indeed non-convex and thus hard to solve. An alternative approach was proposed in \cite{rahimi2009weighted} where one-hidden-layer neural networks are thought as Monte-Carlo approximations of kernel expansions. In particular, if we assume that $\omegab$ is a random variable with a support $\Omega$ and a probability distribution $\tau(\omegab)$, the corresponding kernel can be obtained via \cite{rahimi2008random}
\begin{equation}\label{kernel}
    k(\xb,\xb')=\int_\Omega \phi(\xb,\omegab)\phi(\xb',\omegab)d\tau(\omegab).
\end{equation}
Hence, if $\{\omegab_l\}_{l=1}^M$ are independent samples from $\tau(\omegab)$, the approximated kernel expansion corresponds to \eqref{OHL} and learning $\theta_l$ becomes a convex optimization problem with a modest computational cost. $\{\omegab_l\}_{l=1}^M$ are then called {\it random features} in this model.

One such example is using cosine feature map to approximate a Gaussian kernel $k(\xb,\xb') = \exp{\frac{\norm{\xb-\xb'}_2^2}{2}}$ with unit width. In this case, \eqref{OHL} will be as follows
\begin{equation}\label{RF}
    \fh(\xb) = \sum_{l=1}^M \theta_l\sqrt{2}\cos(\nub_l^\top\xb+b_l),
\end{equation}
where $\{\nub_l\}_{l=1}^M$ come from a multi-variate Gaussian distribution $\Nc(\0,\Ib_d)$ and $\{b_l\}_{l=1}^M$ come from a uniform distribution $\Uc(0,2\pi)$.

\subsection{Local Measurements in Multi-agent Networks}\label{2B}
The proposed scenario in the previous section was centralized in the sense that the estimation task was done only by one agent that has all the data $\{(\xb_j,y_j)\}$. In this section, we propose an iterative distributed scheme where we have a network of $n$ agents, each of which has access to a subset of data. In particular, agent $i\in [n]$ has access to only $m_i$ data points at each iteration.

\begin{assumption}\label{A1}
Without loss of generality, we assume each agent observes the same number of data points at each time, i.e., $m_{1} = m_{2} =\cdots=m_{n} = c$  throughout the paper.
\end{assumption}
This assumption is only for the sake of presentation clarity. Our main results can be extended to the case where different agents have various numbers of measurements.

Now, in the distributed model, the observation matrix $\Hb_{i,t} \in \R^{c\times M}$ at time $t$ will be as follows
\begin{equation}\label{observation}
\Hb_{i,t} = \begin{bmatrix}
\phi(\xb_{1,i,t},\omegab_1)&\ldots&\phi(\xb_{1,i,t},\omegab_M)\\
\ldots&\ldots&\ldots\\
\phi(\xb_{c,i,t},\omegab_1)&\ldots&\phi(\xb_{c,i,t},\omegab_M)
\end{bmatrix},
\end{equation}
with any agent $i\in[n]$ having access to $\{\xb_{j,i,t}\}_{j=1}^{c}$. We then have the following measurement model
\begin{equation*}
\yb_{i,t} = \Hb_{i,t}\thetab+\vb_{i,t},
\end{equation*}
where $\thetab=[\theta_1,\ldots,\theta_M]^\top \in \R^M$ is the {\it unknown} parameter that needs to be learned, and $\vb_{i,t}$ denotes the observation noise at agent $i$. The above local measurement model can be interpreted as iteratively collecting low-dimensional measurements of parameter $\thetab$ at $c$ different locations using $M$ distinct frequencies. 

We follow the general assumptions of zero mean and constant variance on the noise term, i.e., we have $\E[\vb_{i,t}] = \0$ and $\E[\vb_{i,t}\vb_{i,t}^\top] = \sigma^2_{v}\Ib_{c}$. We further denote by $\hat{\thetab}_{i,t}$ the estimate of $\thetab$ for agent $i$ at time $t$.

\begin{assumption}\label{A3}
$\thetab=[\theta_1,\ldots,\theta_M]^\top \in \R^M$ is globally identifiable yet locally unobservable, i.e., the following two properties hold:
\begin{itemize}
    \item Rank of $\Gb_i \triangleq c^{-1}\E[\Hb_{i,t}^\top\Hb_{i,t}]$ is strictly less than $M$.
    \item $\sum_{i=1}^n \Gb_i$ is invertible.
\end{itemize}
Note that $\Gb_i$ is also the kernel matrix formed with random features at agent $i$ where its $pq$-th entry is $g_i(\omegab_{p},\omegab_{q})=\ex{\phi(\cdot,\omegab_p)\phi(\cdot,\omegab_q)}$. We are interchanging the role of random features $\omegab$ and data $\xb$ here since both of them are random samples from probability measures.
\end{assumption}
\begin{assumption}\label{A4}
We assume that the feature map is bounded \cite{rahimi2009weighted} and $\sup_{\xb,\omegab}\{|\phi(\xb,\omegab)|\} \leq \sqrt{2}$. This also suggests a trivial bound where $\norm{\Gb_i}_2\leq\tr{\Gb_i} \leq 2M$.
\end{assumption}

\subsection{Multi-agent Network Model}
The interactions of agents, which in turn defines the network, is captured with the matrix $\Pb$. Formally, we denote by $[\Pb]_{ij}$, the $ij$-th entry of the matrix $\Pb$. When $[\Pb]_{ij}>0$, agent $i$ communicates with agent $j$. We assume that $\Pb$ is symmetric, doubly stochastic with positive diagonal elements. The assumption simply guarantees the information flow in the network. Alternatively, from the technical point of view, we respect the following hypothesis. 
\begin{assumption}(connectivity)\label{A2}
The network is connected, i.e., there is a path from any agent $i\in [n]$ to another agent $j\in [n]\setminus \{i\}$. We further assume that $\Pb = \Ib_{n} - \alpha \Lb$, where $\Lb$ is the Laplacian matrix and $0<\alpha<\text{deg}^{-1}$, where $\text{deg}$ denotes the maximum degree of connectivity in the network.
\end{assumption}

The assumption implies that the Markov chain $\Pb$ is irreducible and aperiodic, thus having a unique stationary distribution, i.e., $\1^\top \Pb=\1^\top$ is the unique (unnormalized) left eigenvector corresponding to $\lambda_1(\Pb)=1$. It also entails that $\lambda_1(\Pb)$ is unique, and the other eigenvalues of $\Pb$ are less than unit in magnitude \cite{horn2012matrix}.

\subsection{Distributed Estimation Update}
To construct an iterative estimate of the parameter $\thetab$, each agent $i\in [n]$ at time $t$ performs the following distributed update
\begin{equation}\label{Estimation}
\begin{aligned}
\hat{\thetab}_{i,t+1} &= \sum_{j=1}^{n}\Pb_{ij}\hat{\thetab}_{j,t}+\alpha\Hb_{i,t}^\top(\yb_{i,t}-\Hb_{i,t}\hat{\thetab}_{i,t}),\\
\end{aligned}
\end{equation}
where $\alpha>0$ is the step size. The update is akin to {\it consensus+innovation} schemes in the distributed estimation literature \cite{khan2010connectivity,kar2012distributed,shahrampour2016distributed}, and we analyze this update in Section \ref{sec:main} in the context of one-hidden-layer neural networks. Intuitively, the first part of the update (consensus) allows agents to keep their estimates close to each other, and the second part (innovation) takes into account the new measurements. 

\section{Main Theoretical Results}\label{sec:main}

In this section, we provide our main theoretical results. We show that the local update \eqref{Estimation} is an asymptotically unbiased estimator of the global parameter $\thetab$. Based on this result, we derive the feasible range for step-size to guarantee convergence. We then prove that the asymptotic second moment of the collective estimation error is bounded. 

\subsection{First Moment}
Let us define the {\it local} error for each agent $i\in[n]$ as
\begin{equation}\label{error}
\eb_{i,t} \triangleq \hat{\thetab}_{i,t} - \thetab.
\end{equation}
Subtracting $\thetab$ from both sides of the local update \eqref{Estimation}, we can write the iterative local error process as follows
\begin{equation}\label{errorupdate}
\eb_{i,t+1} = \sum_{j=1}^{n}\Pb_{ij}\eb_{j,t} - \alpha\Hb_{i,t}^\top\Hb_{i,t}\eb_{i,t}+\alpha\Hb_{i,t}^\top\vb_{i,t}.
\end{equation}
Stacking the local errors in a vector, we denote the {\it global} error by 
\begin{align}\label{eq:errorglobal}
\eb_t\triangleq[\eb_{1,t}^\top,\ldots,\eb_{n,t}^\top]^\top.
\end{align}
We now characterize the global error process with the following proposition.
\begin{proposition}\label{P1}
Given Assumptions \ref{A1}-\ref{A2}, the expected global error can be expressed as an LTI system that takes the form
\begin{equation*}
    \E[\eb_t] = \Qb \E[\eb_{t-1}],
\end{equation*}
where 
\begin{equation}\label{Q}
      \Qb \triangleq \Ib_{Mn} - \alpha \Bb ~~~~~~~ \Bb \triangleq \Lb \otimes \Ib_M + c \Gb,  
\end{equation}
and $\otimes$ denotes the Kronecker product, $\Gb \triangleq \text{diag}[\Gb_1,\ldots,\Gb_n]$ and $\{\Gb_i\}_{i=1}^n$ is defined in Assumption \ref{A3}. The expectation is taken over the stochasticity of $\xb$ and $\vb$. \hfill $\Box$
\end{proposition}

The proof of proposition \ref{P1} is given in the Appendix. It shows that the agents will collectively generate estimates of the parameter $\thetab$ that are asymptotically unbiased as long as the spectral radius of $\Qb$ is less than 1.

\subsection{Step Size Tuning}\label{tuning}
According to Proposition \ref{P1}, a sufficient condition for the convergence of the first moment is that the spectral radius of $\Qb$ should be less than 1. The spectral radius of $\Qb$ is decided by the following two quantities:
\begin{equation}\label{largesteigen}
    \lambda_1(\Qb) = 1-\alpha \lambda_{Mn}(\Bb),
\end{equation}
and
\begin{equation}\label{smallesteigen}
    \lambda_{Mn}(\Qb) = 1-\alpha \lambda_1(\Bb).
\end{equation}
%We can then infer that
%\begin{equation}
 %   \begin{aligned}
  %  \rho(\Qb) = \max(1-\alpha \lambda_{Mn}(\Bb),\alpha \lambda_1(\Bb)-1).
  %  \end{aligned}
%\end{equation}
Now, given the condition for convergence $\rho(\Qb)< 1$, we can derive the feasible range for step size $\alpha$. According to Assumption \ref{A2}, $\1_n$ is the (un-normalized) eigenvector of the matrix $\Lb$ associated with the unique zero eigenvalue $\lambda_n(\Lb)=0$, because $\Lb\1_n = 0$. Therefore, due to Assumption \ref{A3}, $\Gb\1_{Mn} >0$
and $\Bb$ is always positive definite. It is then immediate that $1-\alpha \lambda_{Mn}(\Bb) < 1$. On the other hand, 
\begin{equation}
    \begin{aligned}
        \alpha \lambda_1(\Bb)-1 < 1 \Longleftrightarrow & \alpha < \frac{2}{\lambda_1(\Bb)}.
    \end{aligned}
\end{equation}

In conclusion, a sufficient condition for first moment convergence of global error is $\alpha < 2/\lambda_1(\Bb)$.

\subsection{Asymptotic Second Moment} 
To capture the efficiency of the collective estimation, we should also study the variance of the error, which (asymptotically) amounts to the second moment in view of Proposition \ref{P1}. In the next theorem, we present an asymptotic upper bound on the second moment for a feasible range of step size $\alpha$.

\begin{theorem}\label{T2}
Given Assumptions \ref{A1}-\ref{A2}, the expected second moment of the estimation error is bounded under the following condition. When $\alpha < \frac{2\lambda_{Mn}(\Bb)}{(\lambda_1(\Lb) + 2Mc)^2}$,
\begin{equation*}
\lim\limits_{t \rightarrow \infty}\E[\eb_{t}^\top\eb_{t}] \leq \frac{2\alpha c M n\sigma^2_{v}}{2\lambda_{Mn}(\Bb) - \alpha (\lambda_1(\Lb) + 2Mc)^2}.
\end{equation*}
The expectation is taken over the stochasticity of data $\xb$ and observation noise $\vb$. \hfill $\Box$
\end{theorem}

The proof of theorem \ref{T2} is given in the Appendix. It shows that the (asymptotic) expected second moment of the estimation error is bounded by a finite value that scales linearly with respect to the number of agents $n$ for a certain range of step size $\alpha$.  

\section{Numerical Experiments}\label{sec:simulation}
We now provide empirical evidence in support of our algorithm by applying it to a regression dataset on UCI Machine Learning Repository\footnote{https://archive.ics.uci.edu/ml/datasets/Appliances+energy+prediction}. In this dataset, the input $\xb \in \R^{28}$ includes a number of attributes including temperature in kitchen area, humidity in kitchen area, temperature in living room area, humidity in laundry room area, temperature outside, pressure, etc.. The regression model aims at representing appliances energy use in terms of these features. More details about this dataset can be found in \cite{candanedo2017data} as well as the UCI Machine Learning Repository. We randomly choose 16000 observations out of its 19735 observations for our simulation.

We consider observation matrices $\Hb_{i,t}$ of form \eqref{observation}, where the bases are cosine functions as follows
\begin{equation}
    \phi (\xb,\omegab)=\phi (\xb,\nub,b)=\sqrt{2}\cos(\xb^\top\nub+b),
\end{equation}
as described in Section \ref{2A} where $\{\nub_l\}_{l=1}^M$ come from a multi-variate Gaussian distribution $\Nc (\0,\Ib_d)$ and $\{b_l\}_{l=1}^M$ come from a uniform distribution $\Uc(0,2\pi)$. Without loss of generality, we set $M=5$, i.e., we use five basis functions in the approximation model \eqref{RF}. One can consider other values for $M$ and perform cross-validation to find the best one, but this is outside of the scope of this paper, as our focus is on estimation rather than model selection.

\noindent
{\bf Network Structure:}
We consider a network of $40$ agents. Each agent $i$ has access to observation matrix $\Hb_{i,t}$ with $c=40$ data points at time $t$. Also, each agent $i$ is connected to $4$ agents $i-2,i-1,i+1,i+2$ (with a circular shift for any number outside of the range $[1,40]$). The matrix $\Pb$ is such that agent $i$ is connected to itself with weight $0.84$ and connected to agents $i-2,i-1,i+1,i+2$ with weight $0.04$. According to estimation, the largest eigenvalue of $\Bb$ is $\lambda_1(\Bb)=42$, so according to the step size constraint for the first moment convergence, the feasible step size range is $\alpha < 0.047$. Also, the step size recommendation in Theorem \ref{T2} is $\alpha < 0.0005$, but to achieve a faster convergence, we set $\alpha = 0.04$, which violates the latter condition in theory but works in practice.

\noindent
{\bf Benchmark:}
Since this dataset is real-world and the ground truth value $\thetab$ is unknown, we consider the solution of the centralized problem as the baseline. The local error at time $t$ is then calculated as the difference between local estimates $\hat{\thetab}_{i,t}$ and the centralized estimates as given in \eqref{error}. We run update \eqref{Estimation} for $30000$ iterations such that the process reaches a steady state. To verify our results, we repeat the update process using Monte-Carlo simulations for $100$ times by giving the agents random data points to estimate the expectations.

\begin{figure}[t]
\centering
\includegraphics[trim=0mm 0mm 0mm 0mm,clip, scale=0.4]{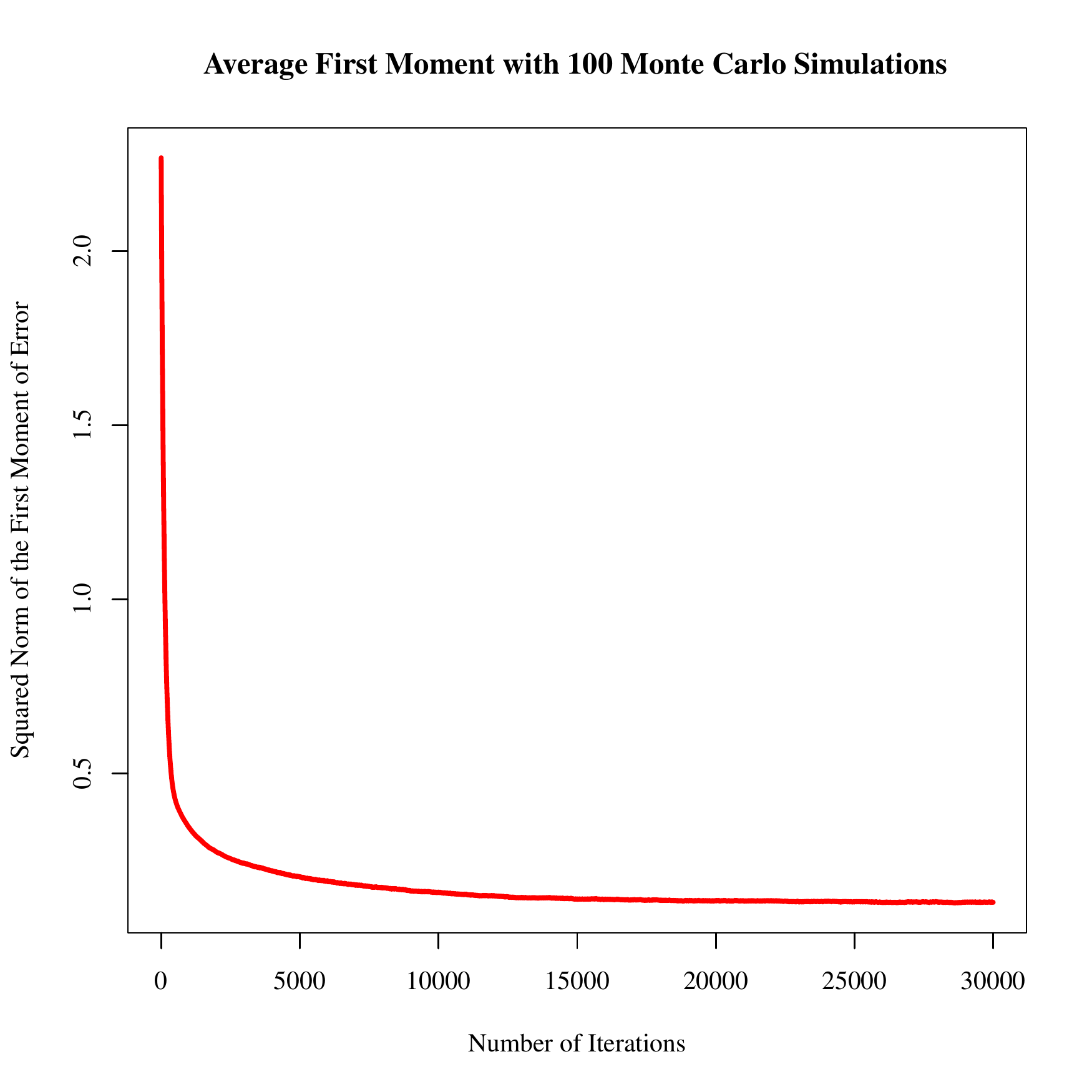}
\caption{The squared norm-$2$ of averaged (over agents) global error converges to zero over 100 Monte-Carlo simulations as the number of iterations increases.}
\label{fig: norm 1}
\end{figure}

\noindent
{\bf Performance:} We visualize the error process in Proposition \ref{P1} by presenting the plot of squared norm-$2$ of the expected global error (averaged over agents), i.e., the squared norm-$2$ of $\E[\eb_t]$ (divided by $40$) given in Proposition \ref{P1} against number of iterations $t$. The vertical axis in Fig. \ref{fig: norm 1} represents the average global error obtained by repeating Monte-Carlo simulations to form an estimate of the expected global error. The horizontal axis shows the number of iterations. By setting the number of Monte-Carlo simulations as $100$, we can expect the squared norm-$2$ of the average global error converging to the squared norm-$2$ of the expected global error in Proposition \ref{P1}. As we can observe, the estimation of the expected global error converges to zero verifying that agents form asymptotically unbiased estimators of the parameter. 

\begin{figure}[t]
\centering
\includegraphics[trim=0mm 0mm 0mm 0mm,clip, scale=0.4]{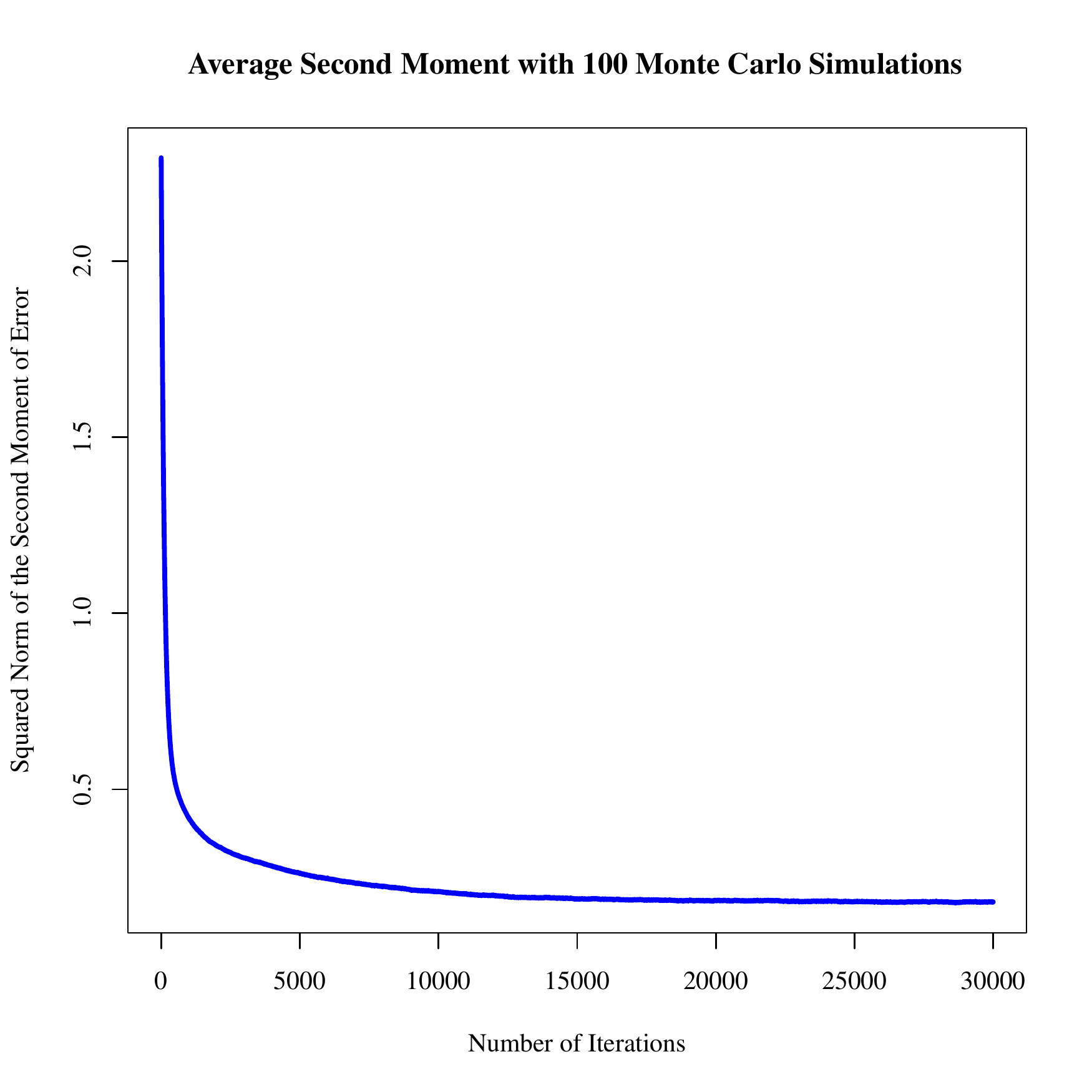}
\caption{The estimates across all agents have a finite variance.}
\label{fig: norm 2}
\end{figure}

We next plot the expected squared norm-$2$ of global error, i.e., $\E[\eb_t^\top \eb_t]$ (divided by $40$) given in Theorem \ref{T2} estimated over $100$ Monte-Carlo simulations.  The vertical axis in Fig. \ref{fig: norm 2} represents the squared norm-$2$ of the global error averaged over Monte-Carlo simulations. The horizontal axis shows the number of iterations. We observe that though the step size does not satisfy the (sufficient) condition in Theorem \ref{T2}, the second moment converges.  

\section{Conclusion}
In this paper, we considered a distributed scheme for parameter estimation in randomized one-hidden-layer neural networks. A network of agents exchange local estimates of the parameter, formed using partial observations, to collaboratively identify the true value of the parameter. Our main contribution is to characterize the behavior of this distributed estimation scheme. We showed that the global estimation error is asymptotically unbiased and its second moment is finite under mild assumptions. Interestingly, our results shed light on the interplay of step size and network structure, which can be used for optimal design in practice. We verified this empirically by applying our method to a real-world data. Future directions include studying the estimation problem when the parameter has some dynamics \cite{shahrampour2013online} or the random frequencies are generated from a time-varying distribution. Due to the non-stationary nature of the problem in these two cases, the theoretical analysis becomes challenging and interesting to explore.

\addtolength{\textheight}{-0cm}   % This command serves to balance the column lengths

%%%%%%%%%%%%%%%%%%%%%%%%%%%%%%%%%%%%%%%%%%%%%%%%%%%%%%%%%%%%%%%%%%%%%%%%%%%%%%%%

%%%%%%%%%%%%%%%%%%%%%%%%%%%%%%%%%%%%%%%%%%%%%%%%%%%%%%%%%%%%%%%%%%%%%%%%%%%%%%%%

%%%%%%%%%%%%%%%%%%%%%%%%%%%%%%%%%%%%%%%%%%%%%%%%%%%%%%%%%%%%%%%%%%%%%%%%%%%%%%%%
\section*{Appendix}
For presentation clarity, we use the following definitions in the proofs:
\begin{align*}
\Bb_t&\triangleq \Lb \otimes \Ib_M + \text{diag}[\Hb_{1,t}^\top\Hb_{1,t},\ldots,\Hb_{n,t}^\top\Hb_{n,t}]\\
\Eb_{i,t}&\triangleq \Hb_{i,t}^\top\vb_{i,t}\\
\Eb_{t}&\triangleq[\Eb_{1,t}^\top,\ldots,\Eb_{n,t}^\top]^\top.   \numberthis\label{eq:defs}
\end{align*}

\subsection{Proof of Proposition \ref{P1}}
%To prove Proposition \ref{P1}, we first need to show that
%\begin{equation}\label{eq:hth}
%\E[\Hb_{i,t}^\top\Hb_{i,t}] = c\Gb_i,  
%\end{equation}
%for any $i\in[n]$, where $\Gb_i$ is the kernel matrix formed by random features. 

%Notice that given the observation model \eqref{observation}, the $pq$-th entry of the matrix $\Hb_{i,t}^\top\Hb_{i,t}$ can be written as
%\begin{equation}
%[\Hb_{i,t}^\top\Hb_{i,t}]_{pq} = \sum_{j=1}^{c}\phi(\xb_{j,i,t},\omegab_p)\phi(\xb_{j,i,t},\omegab_q).
%\end{equation}
%Then each entry is $[\E[\Hb_{i,t}^\top\Hb_{i,t}]]_{pq} = cg(\omegab_p,\omegab_q)$, since for any $\omegab \in \Omega^d$ we have
%\begin{equation*}
        %\E[\phi(\xb,\omegab_p)\phi(\xb,\omegab_p)] = g(\omegab_p,\omegab_q).
%\end{equation*}
Notice that $\E[\Hb_{i,t}^\top\Hb_{i,t}]=c\Gb_i$, entailing that
\begin{align}\label{eq:ddd}
\E[\Bb_t] = \Lb \otimes \Ib_M + c\text{diag}[\Gb_1,\ldots,\Gb_n] = \Bb,
\end{align}
in view of \eqref{eq:defs}.
Following the lines of the proof of Lemma 1 in \cite{shahrampour2016distributed}, the error process can be expressed as the following 
\begin{align}\label{eq:errorg}
\eb_{t+1}=\Qb'_t\eb_t+\alpha\Eb_t,
\end{align}
where
\begin{equation}\label{eq:qprime}
    \Qb'_t = \Ib_{Mn} - \alpha\Bb_t.
\end{equation}
Taking expectation over data on both sides and noting \eqref{eq:ddd}, we have 
\begin{equation*}
\Qb \triangleq \E[\Qb'_t]= \Ib_{Mn} - \alpha\E[\Bb_t]=\Ib_{Mn} - \alpha\Bb.
\end{equation*}
Recalling \eqref{eq:defs}, we can also immediately see from the zero-mean assumption on the noise that $\E[\Eb_{i,t}]=\0$ for every $i\in [n]$. Combining this with above and returning to \eqref{eq:errorg} will finish the proof of Proposition \ref{P1}.

\subsection{Proof of Theorem \ref{T2}}
To prove Theorem \ref{T2}, we first need to show a recursive relationship for the error process based on \eqref{eq:errorg} where
\begin{equation}\label{second moment update}
\begin{aligned}
\ex{\eb_{t+1}^\top\eb_{t+1}}&=\ex{(\Qb'_t\eb_t+\alpha\Eb_t)^\top(\Qb'_t\eb_t+\alpha\Eb_t)}\\
&= \ex{\eb_t^\top {\Qb'_t}^\top\Qb'_t\eb_t} + \alpha^2\ex{\Eb_t^\top\Eb_t}\\
&\leq \rho\left(\ex{{\Qb'_t}^\top\Qb'_t}\right)\ex{\eb_t^\top\eb_t} + \alpha^2\ex{\Eb_t^\top\Eb_t}\\
&=\lambda_1\left(\ex{{\Qb'_t}^\top\Qb'_t}\right)\ex{\eb_t^\top\eb_t} + \alpha^2\ex{\Eb_t^\top\Eb_t},
\end{aligned}
\end{equation}
where we used the fact $\E[\vb_{i,t}]=\0$, resulting in zero cross-terms in the second line. To further bound $\lambda_1(\E[{\Qb'_t}^\top\Qb'_t])$, let us recall \eqref{eq:qprime}, we have that  
\begin{align*}
\E\Big[{\Qb'_t}^\top\Qb'_t\Big]&= \E\Big[\Ib_{Mn} - 2\alpha \Bb_t + \alpha^2\Bb_t^2\Big]\\
&= \Ib_{Mn} - 2\alpha \Bb + \alpha^2 \E\Big[\Bb_t^2\Big].
\end{align*}
Now, observe that $\lambda_1(\Bb_t) \leq \lambda_1(\Lb) + 2Mc$ due to Assumption \ref{A4}. We can bound the spectral radius of the above matrix as
\begin{equation}\label{phi_a}
    \begin{aligned}
        \lambda_1(\E[{\Qb'_t}^\top\Qb'_t]) &\leq 1 - 2\alpha\lambda_{Mn}(\Bb) + \alpha^2 (\lambda_1(\Lb) + 2Mc)^2.
    \end{aligned}
\end{equation}

%We denote the above upper bound matrix by $\Sb$. Now, let us pre/post multiply the above matrix by vector of all ones $\1$ with dimension $Mn$, we can get the following

%\begin{equation}
%    \begin{aligned}
%        &\1^\top \Sb \1\\
 %       =&\1^\top(\Pb \otimes \Ib_M)^2\1 - \alpha c \1^\top (\Pb \otimes \Ib_M)\Gb\1 \\
  %      &- \alpha c \1^\top\Gb(\Pb \otimes \Ib_M)\1 + 4\alpha^2 M^2 c^2 \1^\top\1\\
   %     =& \1^\top\1 - 2\alpha c \1^\top\Gb\1 + 4\alpha^2 M^2 c^2 \1^\top\1\\
    %    \Rightarrow& \frac{\1^\top\Sb\1}{\1^\top\1} = 1+4\alpha^2 M^2 c^2 - 2\alpha c \frac{\1^\top\Gb\1}{\1^\top\1}.
    %\end{aligned}
%\end{equation}
%Therefore for an arbitrarily small $\alpha$, the following bound holds
%\begin{equation}\label{phi_a}
%    \lambda_1(\E[{\Qb'_t}^\top\Qb'_t]) \leq 1+4\alpha^2 M^2 c^2 -2\alpha c \lambda_M(\overline{\Gb}).
%\end{equation}

Recalling \eqref{eq:defs}, we can then bound the additive term in the recursive relation \eqref{second moment update} as follows
\begin{equation}\label{phi_b}
\begin{aligned}
\alpha^2 \E\bigg[\Eb_{t}^\top\Eb_{t}\bigg]= &\alpha^2 \E\bigg[\sum_{i=1}^n \Eb_{i,t}^\top\Eb_{i,t}\bigg]\\
= &\alpha^2 \E\bigg[\sum_{i=1}^n \vb_{i,t}^\top \Hb_{i,t}\Hb_{i,t}^\top\vb_{i,t}\bigg]\\
= &\alpha^2  \sum_{i=1}^n \tr{\E\big[\Hb_{i,t}\Hb_{i,t}^\top\big]\E\big[\vb_{i,t}\vb_{i,t}^\top\big]}\\
= &\alpha^2  \sum_{i=1}^n \tr{\E\big[\Hb_{i,t}^\top\Hb_{i,t}\big]}\sigma^2_{v}\\
= &\alpha^2 c \sum_{i=1}^n \tr{\Gb_i}\sigma^2_{v} \leq 2\alpha^2 c M n\sigma^2_{v}.
\end{aligned}
\end{equation}
Letting 
\begin{align*}
\Phi_a&\triangleq 1 - 2\alpha\lambda_{Mn}(\Bb) + \alpha^2 (\lambda_1(\Lb) + 2Mc)^2\\
\Phi_b&\triangleq 2\alpha^2 c M n\sigma^2_{v}, \numberthis\label{phis}
\end{align*}
and using \eqref{phi_a} and \eqref{phi_b}, we can re-write the recursive relation in \eqref{second moment update} as
\begin{equation}\label{recursive}
    \E[\eb_{t+1}^\top\eb_{t+1}] \leq \Phi_a \E[\eb_{t}^\top\eb_{t}] + \Phi_b.
\end{equation}

We can find the feasible range of $\alpha$ through the inequality $\Phi_a < 1$ which ensures that the recursive process \eqref{recursive} will converge.

\begin{equation}
    \begin{aligned}
        \vphantom{\alpha < \frac{2\lambda_{Mn}(\Bb)}{(\lambda_1(\Lb) + 2Mc)^2}}\Phi_a < 1 \Longleftrightarrow & 1 - 2\alpha\lambda_{Mn}(\Bb) + \alpha^2 (\lambda_1(\Lb) + 2Mc)^2 < 1\\
        \vphantom{\alpha < \frac{2\lambda_{Mn}(\Bb)}{(\lambda_1(\Lb) + 2Mc)^2}}\Longleftrightarrow & \alpha^2 (\lambda_1(\Lb) + 2Mc)^2 < 2\alpha\lambda_{Mn}(\Bb)\\
        \Longleftrightarrow & \alpha < \frac{2\lambda_{Mn}(\Bb)}{(\lambda_1(\Lb) + 2Mc)^2}.
    \end{aligned}
\end{equation}

%First, we have the following fact
%\begin{equation*}
    %\resizebox{1\hsize}{!}{%
   %     $\lambda_1((\Pb \otimes \Ib_M-\alpha c\Gb)^2) = \max %\{(1-\alpha c \lambda_{M}(\overline{\Gb}))^2,(\alpha c %\lambda_1(\Gb) - \lambda_n(\Pb))^2\}.$%
   %     }
%\end{equation*}

%One can show that $\lambda_1((\Pb \otimes \Ib_M-\alpha c\Gb)^2) = (1-\alpha c \lambda_{M}(\overline{\Gb}))^2$ when $\alpha \leq \frac{1+\lambda_n(\Pb)}{c(\lambda_{M}(\overline{\Gb}) + \lambda_1(\Gb))}$ and $\lambda_1((\Pb \otimes \Ib_M-\alpha c\Gb)^2) = (\alpha c \lambda_1(\Gb) - \lambda_n(\Pb))^2$ otherwise. 

%We present the result for the case when $\alpha \leq \frac{1+\lambda_n(\Pb)}{c(\lambda_{M}(\overline{\Gb}) + \lambda_1(\Gb))}$, we have the following
%\begin{equation*}
%    \begin{aligned}
%        &\Phi_a <1\\
%        \Longleftrightarrow & (1-\alpha c \lambda_{M}(\overline{\Gb}))^2 + 4\alpha^2c^2M^2-\alpha^2 c^2\lambda_{min}^2(\Gb) <1\\
%        \Longleftrightarrow & \alpha^2c^2\lambda_{M}^2(\overline{\Gb}) + 4\alpha^2c^2M^2 -\alpha^2 c^2\lambda_{min}^2(\Gb)   < 2\alpha c \lambda_{M}(\overline{\Gb})\\
%        \Longleftrightarrow & \alpha < \frac{2\lambda_{M}(\overline{\Gb})}{c\lambda_{M}^2(\overline{\Gb}) + 4cM^2 -  c\lambda_{min}^2(\Gb)}.
%    \end{aligned}
%\end{equation*}

Therefore, given $\alpha < \frac{2\lambda_{Mn}(\Bb)}{(\lambda_1(\Lb) + 2Mc)^2}$, we have that
\begin{equation*}
\begin{aligned}
    \vphantom{\frac{\Phi_b(1-\Phi_a^t)}{1-\Phi_a}}\E[\eb_{t+1}^\top\eb_{t+1}] &\leq \Phi_a \E[\eb_{t}^\top\eb_{t}] + \Phi_b\\
    \vphantom{\frac{\Phi_b(1-\Phi_a^t)}{1-\Phi_a}}&\leq \Phi_a^t\E[\eb_1^\top\eb_1] + \Phi_b(\Phi_a^{t-1}+...+\Phi_a+1)\\
    \vphantom{\frac{\Phi_b(1-\Phi_a^t)}{1-\Phi_a}}&= \Phi_a^t\E[\eb_1^\top\eb_1] + \frac{\Phi_b(1-\Phi_a^t)}{1-\Phi_a}.
\end{aligned}
\end{equation*}
This upper bound will converge to $\frac{\Phi_b}{1-\Phi_a}$ as $t \rightarrow 0$, and noting definitions of $\Phi_a$ and $\Phi_b$ in \eqref{phis}, we derive the upper bound in the statement of Theorem \ref{T2}.
%\addtolength{\textheight}{-12cm}

%\subsection{Statement and Proof of Lemma \ref{L3}}
%\begin{lemma}\label{L3}
%Under same assumptions as Theorem \ref{T2}, 
%\begin{equation}\label{partial order}
 %\E[\Bb_t^2] \preceq 4 c^2 M^2\Ib_{Mn}, 
%\end{equation}
%where $\Bb_t$ is defined in \eqref{eq:defs}. \hfill $\Box$
%\end{lemma}
%\begin{proof}
%In the proof, we omit the time index $t$ and agent index $i$ for presentation clarity, i.e., we denote $\Hb_{i,t}$ by $\Hb$, $\xb_{k,i,t}$ by $\xb_{k}$ for any $k\in[c]$. We will show that $\E[\Hb^\top\Hb\Hb^\top\Hb]$ is upper bounded by $4c^2M^2$.

%Let us start by observing that the $p$-th diagonal entry of the matrix $\Hb^\top\Hb\Hb^\top\Hb$ (for any agent) can be written as 
%\begin{equation}\label{sum}
% \resizebox{1\hsize}{!}{%
%$\sum\limits_{j=1}^M(\sum\limits_{k=1}^{c} \phi(\xb_k,\omegab_p)\phi(\xb_{k},\omegab_j) \sum\limits_{k'=1}^{c} \phi(\xb_{k'},\omegab_j)\phi(\xb_{k'},\omegab_p)),
%$
%
%}
%\end{equation}
%We now consider a single term in the previous summation:
%\begin{equation}\label{term}
%\phi(\xb_k,\omegab_p)\phi(\xb_k,\omegab_j)\phi(\xb_{k'},\omegab_j)\phi(\xb_{k'},\omegab_q).
%\end{equation}
%Recall Assumption \ref{A4} that $\phi(\xb,\omegab) \leq \sqrt{2}$, it is immediate that term \eqref{term} is upper bounded by $4$. Since there are $Mc^2$ equivalent terms \eqref{term} in \eqref{sum}, each diagonal entry of $\E[\Hb^\top\Hb\Hb^\top\Hb]$ is upper bounded by $4c^2M$, furthermore, we can bound its largest eigenvalue by its trace which is $4c^2M^2$. Lemma \ref{L3} is proved.
%\end{proof}

\section*{Acknowledgments}
We gratefully acknowledge the support of  NSF ECCS-1933878 Award as well as Texas A\&M Triads for Transformation (T3) Program. 

\bibliographystyle{IEEEtran}
\bibliography{References,shahin_RF}

%%%%%%%%%%%%%%%%%%%%%%%%%%%%%%%%%%%%%%%%%%%%%%%%%%%%%%%%%%%%%%%%%%%%%%%%%%%%%%%%

\end{document}